\documentclass[
    ,final            
  ]
  {aipproc}
\def\pacs#1{\vspace{10pt} \hspace{0.33cm} \rm PACS numbers: #1 \par \vspace{10pt}}
\layoutstyle{6x9}

\usepackage{amssymb,amsmath}
\usepackage{booktabs}
\usepackage{subfigure}

\makeatletter
\newcommand\setcaptype[1]{\def\@captype{#1}}
\makeatother

\begin{document}

\title{Systematic analysis of transverse momentum distribution and non-extensive thermodynamics theory}


\author{I. Sena}{
  address={Instituto de Fisica -  Universidade de Sao Paulo}
}

\author{A. Deppman}{
  address={Instituto de Fisica -  Universidade de Sao Paulo}
}

\begin{abstract} 
A systematic analysis of transverse momentum distribution of hadrons produced in ultra-relativistic $p+p$ and $A+A$ collisions is presented. We investigate the effective temperature and the entropic parameter from the non-extensive thermodynamic theory of strong interaction. We conclude that the existence of a limiting effective temperature and of a limiting entropic parameter is in accordance with experimental data.
\end{abstract}

\maketitle

General aspects of strong interactions up to center-of-mass energies $\sqrt{s}\sim$ 10 GeV are well understood in terms of a self-consistent theory based on the Boltzmann-Gibbs statistics \cite{Hagedorn}. Hagedorn's theory establishes a connection between the mass spectrum of highly excited hadrons and the density of states for fireballs, and provides correct descriptions for transverse momentum distributions and multiplicities of secondaries.

For $\sqrt{s}\geq$ 10 GeV, however, theoretical and experimental results diverge. A generalized formalism was proposed taking into account the non-extensive thermodynamics~\cite{Bediaga, Beck}. This generalization recovered the agreement between theory and experiment.

Recently it was shown~\cite{Deppman} that a self-consistent theory for hadronic systems based on the non-extensive thermodynamics exists if, for $x \rightarrow \infty$,
\begin{eqnarray}
\rho(x) \rightarrow x^{-5/2} e_q^{\beta_o x}
\end{eqnarray}
and
\begin{eqnarray}
\sigma(x) \rightarrow b x^{a} e_q^{\beta_o x}\,,
\end{eqnarray}
where $\rho(m)$ is the mass spectrum of hadrons and $\sigma(E)$ is the density of states with energy $E$ for hadronic systems, and $a$ is given by
\begin{eqnarray}
a=\frac{\gamma V_o}{2\pi^2 \beta_o^ {3/2}}\,,
\end{eqnarray} 
with $\gamma$ being a constant and $V_o$ being the interaction volume. An important consequence of these constraints is the existence of a limiting effective temperature, $T_o=1/\beta_o$, and of a limiting entropic parameter, $q_o$.

In this work we perform a systematic analysis of experimental data on hadronic collisions to verify if a limiting temperature is consistent with the observed results. 
The set of experimental data used in the present analysis is summarized in Table \ref{tab:Expe} for $p+p$ collisions and Table \ref{tab:dadosIonsPesados} for heavy ions collisions.

    \begin{table}[!h]
    \centering
    \caption{Set of experimental data for $p+p$ collisions.}
    \begin{tabular}{cccc}\toprule
    Experiment                             &     Energy (GeV)              &	$|\eta|$     \\      \midrule
    CMS (LHC)                               &     7000 \cite{CMS7TeV}           &	$|\eta| < 2,4$     \\
    CMS (LHC)                               &     2360 \cite{CMS7TeV}           &	$|\eta| < 2,4$     \\
    CMS (LHC)                               &     900  \cite{CMS7TeV}           &	$|\eta| < 2,4$     \\
    ALICE (LHC)                             &     900  \cite{ALICE900GeV}       &	$|\eta| < 0,8$     \\
    ATLAS (LHC)                             &     900  \cite{ATLAS900GeV}       &	$|\eta| < 2,5$     \\
    RHIC (BNL)                              &     200  \cite{RHIC-STAR-AuAu}    &	$ 3,3 < \eta < 5,0$     \\       \bottomrule
    \end{tabular}
    \label{tab:Expe}
    \end{table}

\begin{table}[!h]
    \centering
    \caption{Set of experimental data for heavy ions collisions.}
    \label{tab:dadosIonsPesados}

      \begin{tabular}{l|c|c|c}\toprule
      Kind of collision&\multicolumn{2}{c|}{Au+Au}    &  Cu+Cu    \\ \midrule
      Energy (GeV)           &  62,4 \cite{Au-62.4}    &  200  \cite{Au-200}             &  200 \cite{Cu-200}     \\ \hline
      Part. produced   &  $\dfrac{h^{+}+h^{-}}{2}$      &  $\pi^{+}$              &  $\dfrac{h^{+}+h^{-}}{2}$      \\ \hline
      $|\eta|$   &  $ 0,2 < \eta < 1,4$      &  $|\eta| < 0,5$              &  $0,2 < \eta < 1,4$      \\ \hline
                 &  0-6      &  0-12              &  0-6      \\
                 &  6-15     &  10-20             &  6-15     \\
                 &  15-25    &  20-40             &  15-25    \\
      Centrality (\%)           &  25-35    &  40-60             &  25-35    \\
                 &  35-45    &  40-80             &  35-45    \\
                 &  45-50    &  60-80             &  45-50    \\
                 &         &  0-80 (\emph{Minimum bias})&          \\ \bottomrule
        \end{tabular}
\end{table}

Our method is to check the existence of a limiting temperature through the study of the transverse momentum ($p_t$) distribution of secondaries produced in the reactions listed above. According to the non-extensive formalism proposed in Refs.~\cite{Bediaga, Beck}, the $p_t$-distribution is given by
\begin{eqnarray}
\frac{1}{\sigma}\frac{d\sigma}{d p_{\perp}}=c[2(q-1)]^{-1/2}B\bigg(\frac{1}{2},\frac{q}{q-1}-\frac{1}{2}\bigg)u^{3/2}[1+(q-1)u]^{-\frac{q}{q-1}+\frac{1}{2}}\,,
\label{pt}
\end{eqnarray}
where $u=\beta_o p_{\perp}$, and $B(x,y)$ is the Beta-function. The dependence of the distribution on the entropic factor, $q$, and on the temperature, given by $\beta = 1/T$, enable us to obtain both parameters by fitting the above expression to experimental data. This procedure has been already used by many authors~\cite{Bediaga, Beck, Chinellato2010}.

Applying this method to $p+p$ collisions we get nice fittings of Eq.~\ref{pt} to the experimental $p_t$-distributions, obtaining the results for $T$ and $q$ that are shown in Fig~\ref{fig:p+p_qANDt_H-T_-TodosPt}. We observe that the temperature varies inside a relatively narrow range between 70 MeV and 90 MeV for all collisions with center-of-mass energy from 0.2 TeV up to 7 TeV. Also, for energies above 1 TeV the temperature can be considered constant with $T\sim$ 73 MeV. Even in the case of the parameter $q$ the variations are relatively small in the energy range studied.
 
   \begin{ltxfigure}[h]
\setcaptype{figure} 
       \centering
       \subfigure[]
                   {\label{fig:T_H-T-p+p}\includegraphics[width=7.cm]{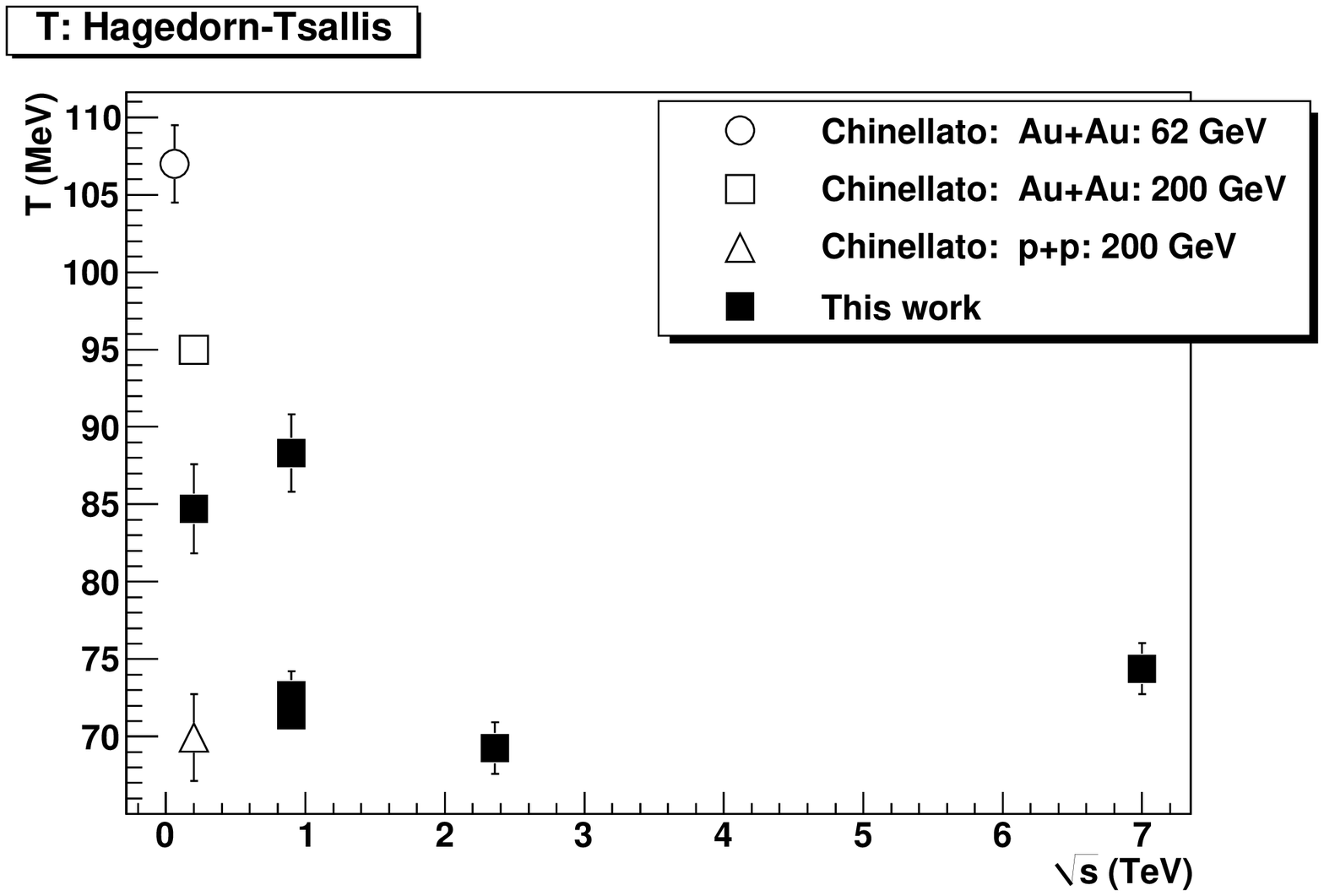}}
       \quad
       \subfigure[]
                   {\label{fig:q_H-T-p+p}\includegraphics[width=7.cm]{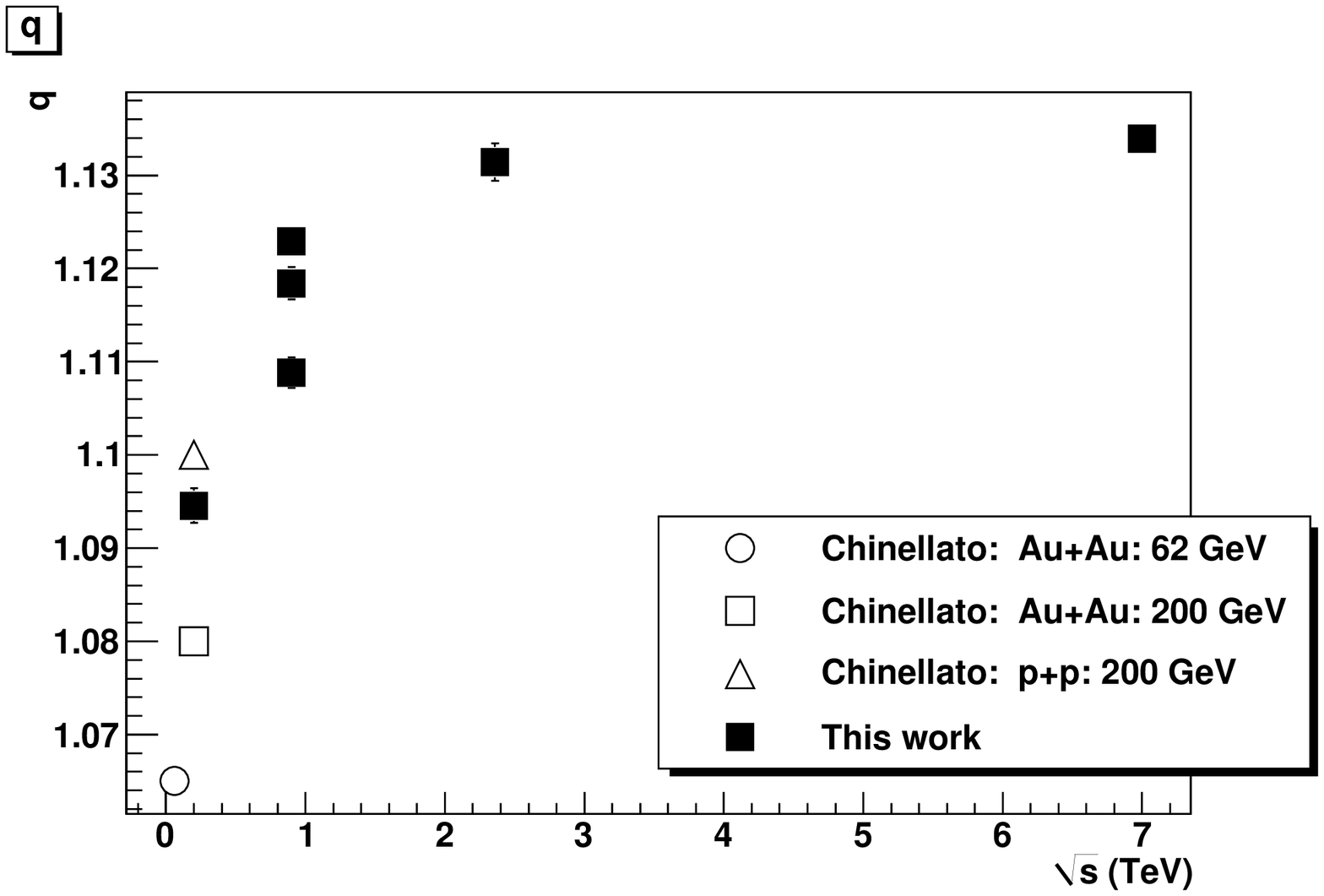}}
          \caption{The values for the parameters corresponding to the best fitted curve to $p_t$-distributions. (a) $T$ as a function of center-of-mass energy. (b) $q$ as a function of center-of-mass energy. Values from Literature are used for comparison.}
       \label{fig:p+p_qANDt_H-T_-TodosPt}
   \end{ltxfigure}

\begin{figure}[!h]
       \centering
                   {\label{fig:T_H-T-p+p}\includegraphics[width=10.cm]{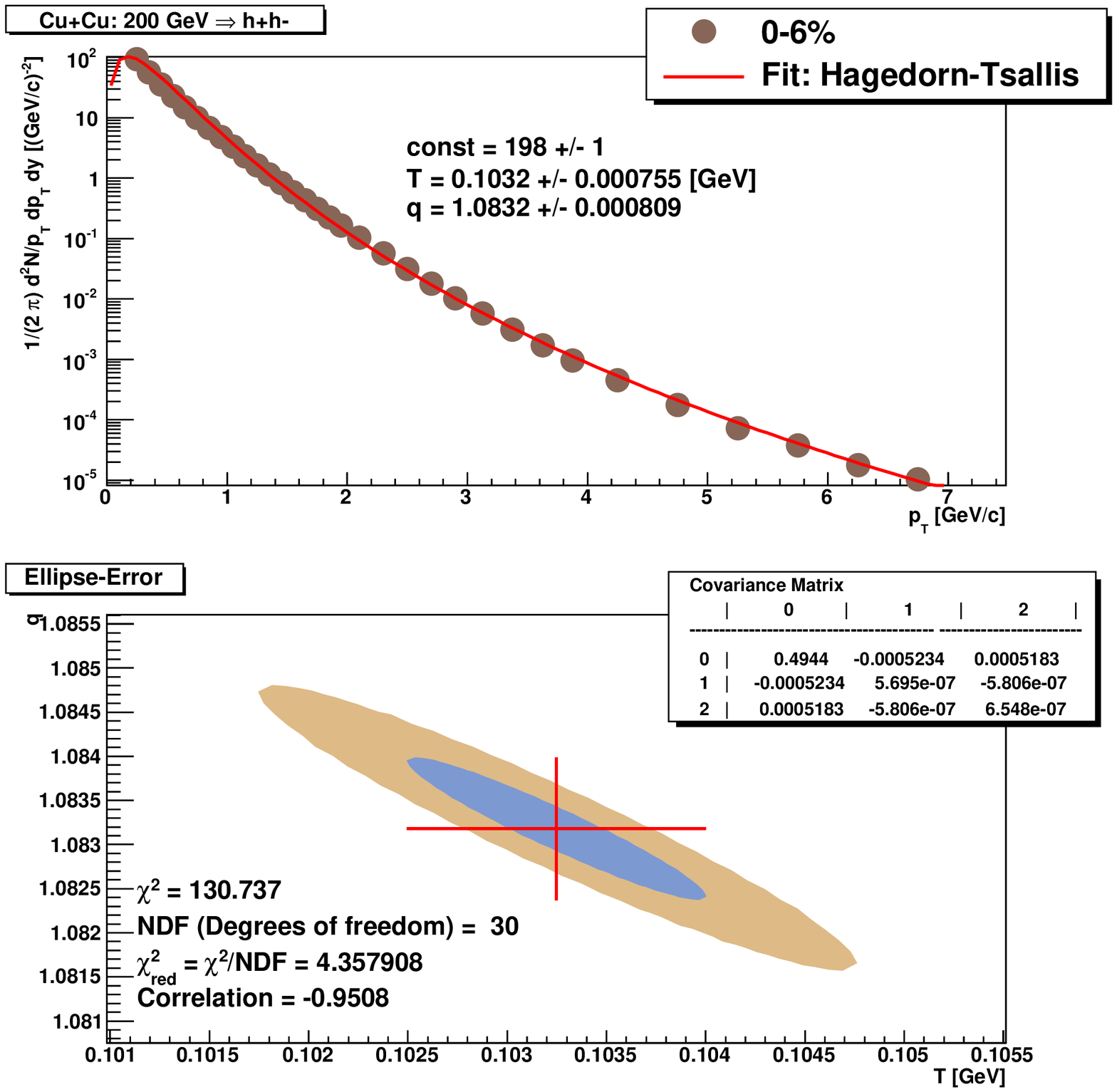}}
          \caption{(a) Best fitted curve to $p_t$-distribution for Cu-Cu collision at $\sqrt{s} = 200$ MeV for centrality in the range 0-6\%. (b) Analysis of the correlation between the parameters $T$ and $q$.}
      \label{fig:Cu+Cu_200_TodosPt_0-6-Estatistica}
      \end{figure}

It is important to notice that $T$ and $q$ in Eq.~\ref{pt} are not completely independent. There is a strong correlation between the two parameters, as can be observed in Fig.~\ref{fig:Cu+Cu_200_TodosPt_0-6-Estatistica}. This correlation can be related to temperature fluctuations of the hadronic system~\cite{Wilk2007, Wilk2009}. In Fig.~\ref{fig:Cu+Cu_200_TodosPt_0-6-Estatistica}(a) we show a typical fit to $p_t$-distribution data, and in Fig.~\ref{fig:Cu+Cu_200_TodosPt_0-6-Estatistica}(b) we show the 2D-plot of  $\chi^2$ distribution for the fit shown in Fig.~\ref{fig:Cu+Cu_200_TodosPt_0-6-Estatistica}(a), where we clearly see an ellipsis which evidences the correlation between the two parameters. Due to their correlation, as we vary $T$ and $q$ simultaneously along the line corresponding to the principal axe of the ellipsis, the $\chi^2$ remains practically unchanged. This means that it is possible to obtain good $\chi^2$ with pairs ($T$, $q$) near the optimum point found in the fitting, indicated by a cross in Figure~\ref{fig:Cu+Cu_200_TodosPt_0-6-Estatistica}(b).

To avoid this correlation we now slightly modify the fitting procedure by adopting a fixed value for $T$, as suggested by the self-consistency principle~\cite{Deppman} and by the results shown in Fig.~\ref{fig:p+p_qANDt_H-T_-TodosPt}. We used different values for $T$ from 70 MeV up to 100 MeV. Also we extended the analysis to $A+A$ collisions, in this case observing the behaviour of the fitted curves at different centralities. Typical results are shown in Fig.~\ref{fig:Au_62.4_Tconst} for $T =$ 90 MeV. As expected we observe good fittings of Eq.~\ref{pt} to the data in all cases, now with only $q$ being a free parameter. Similar results are obtained for different values of $T$ in the range considered, although the quality of the fittings decreases as the temperature departs from values close to $T \sim$ 90 MeV.


      \begin{ltxfigure}[h]
\setcaptype{figure} 
          \centering
          \subfigure[p+p]
                      {\includegraphics[width=7cm]{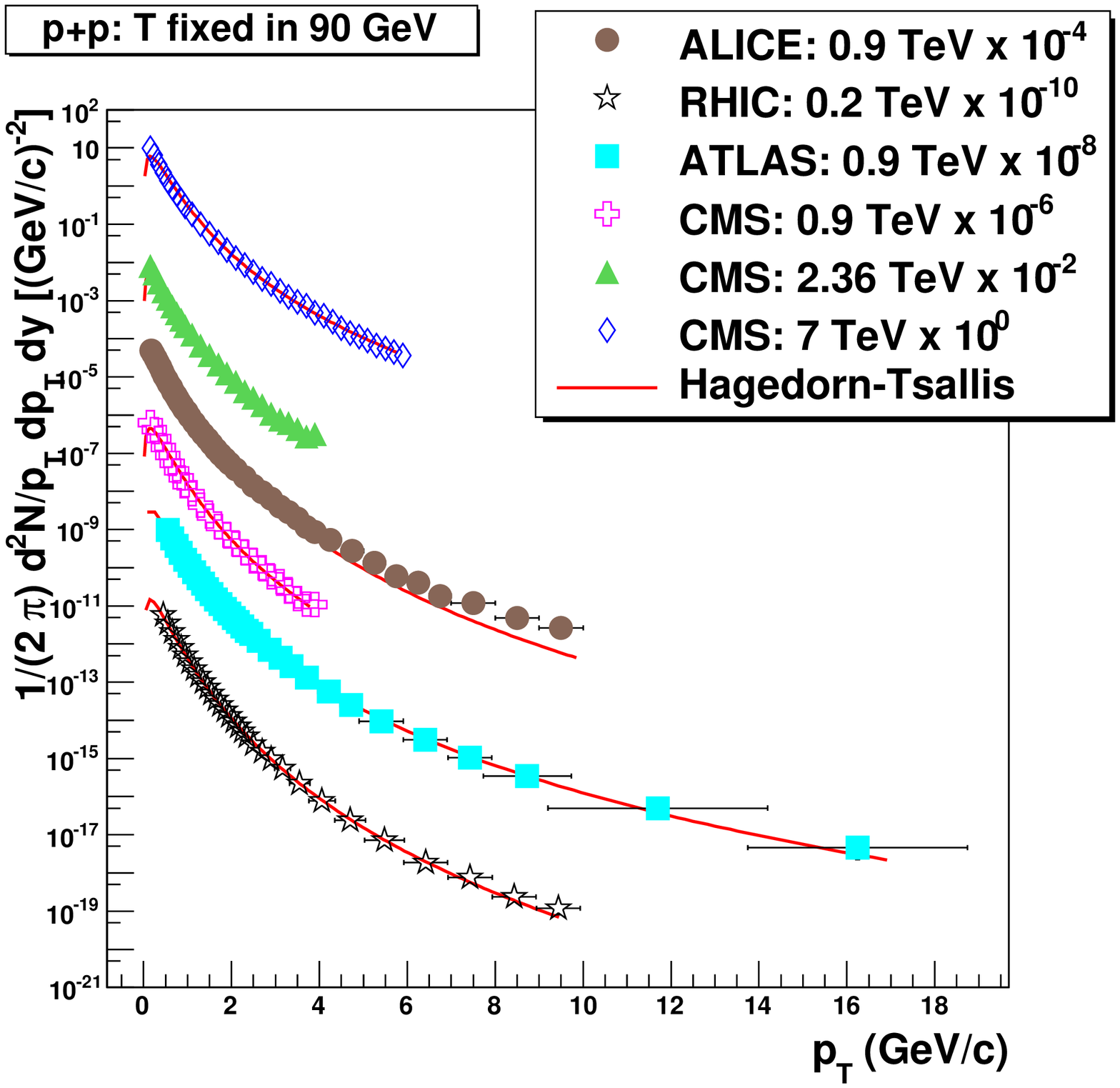}}
          \subfigure[Au+Au: 62.4 GeV]
                      {\includegraphics[width=7cm]{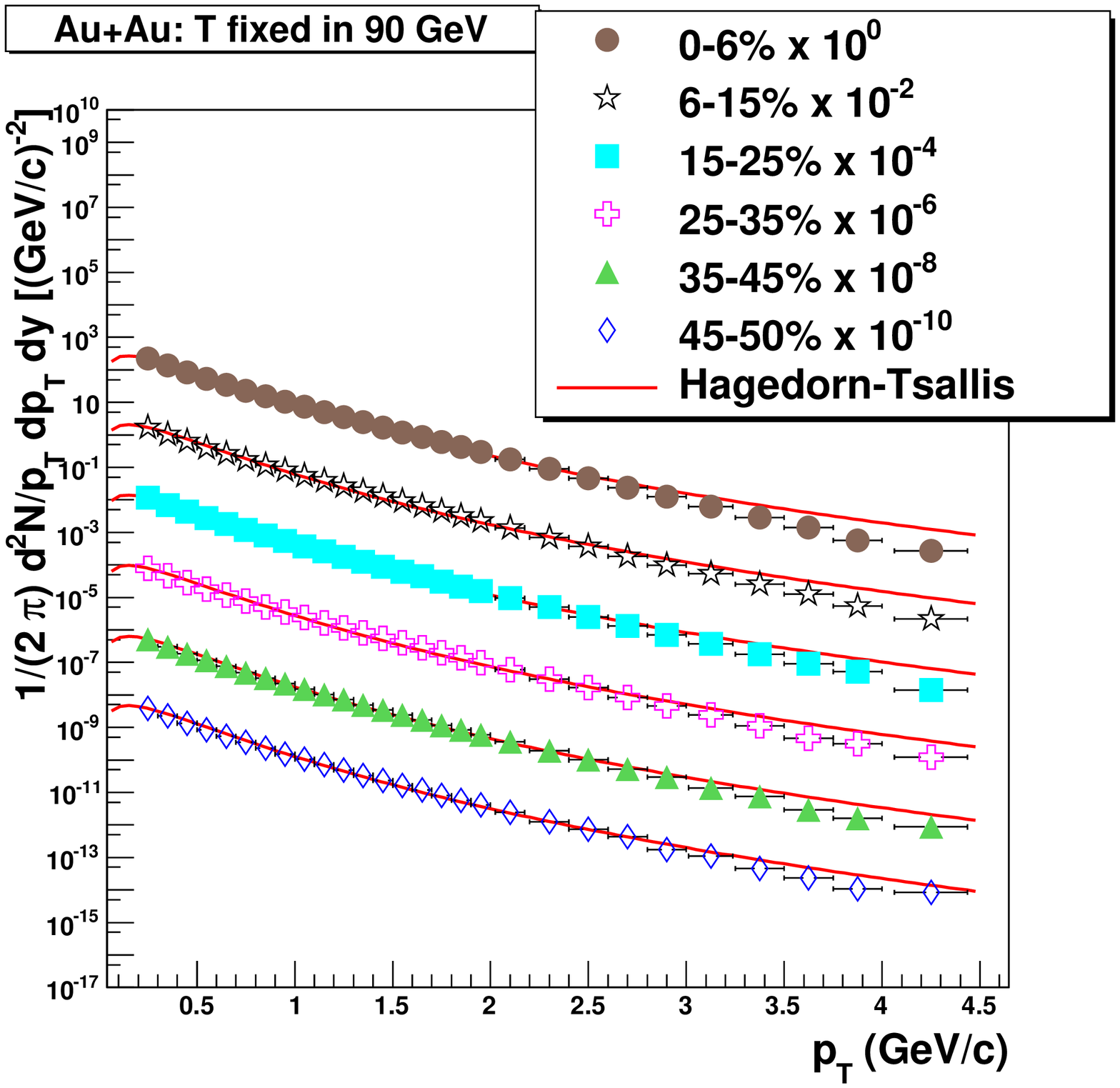}}
          \subfigure[Au+Au: 200 GeV]
                      {\includegraphics[width=7cm]{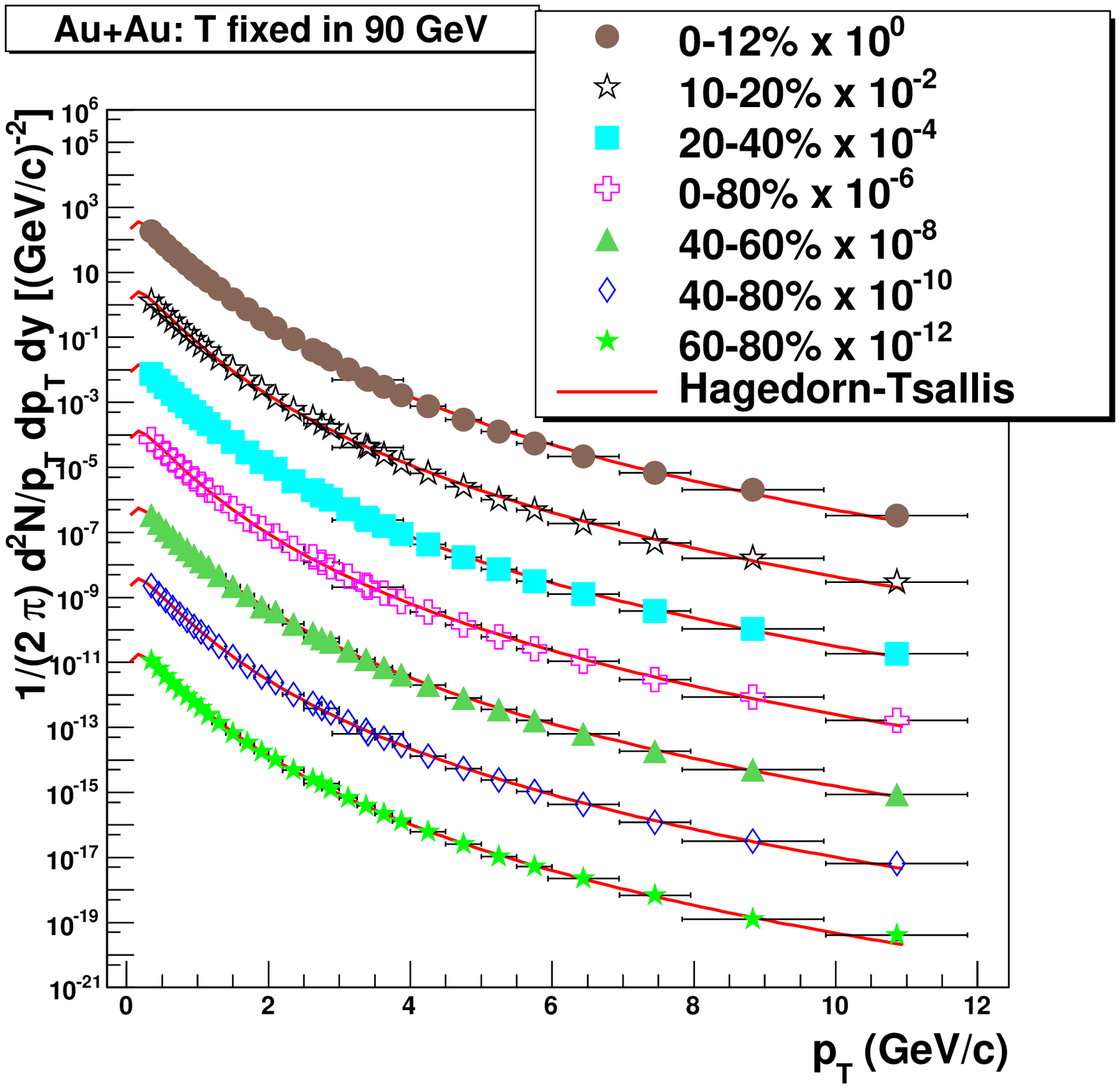}}
          \subfigure[Cu+Cu: 200 GeV]
                      {\includegraphics[width=7cm]{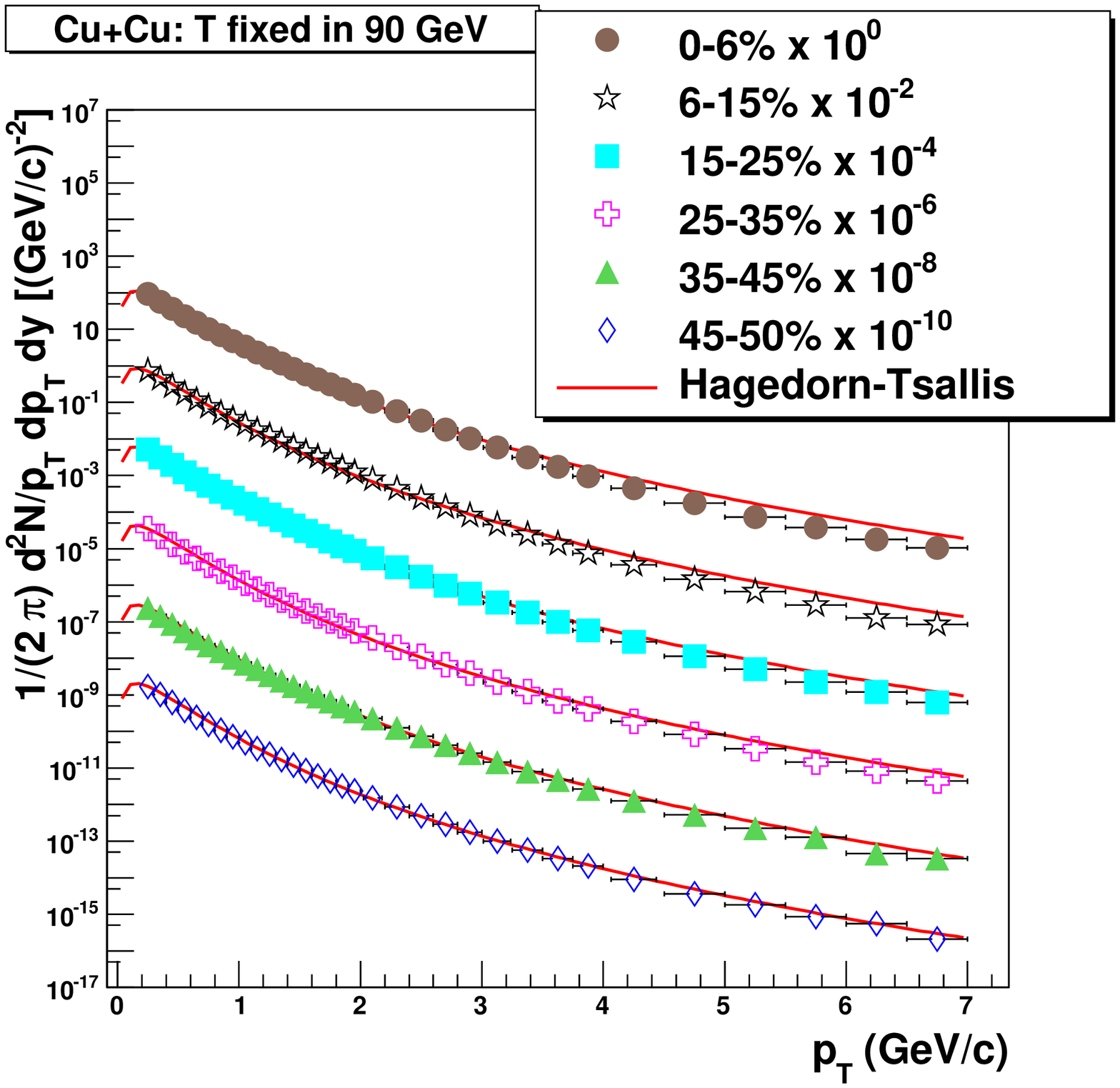}}
   \caption{Typical results for fitting of Eq.~\ref{pt} to experimental data with only $q$ as a free parameter. In this case, $T =$ 90 MeV was adopted.}
          \label{fig:Au_62.4_Tconst}
      \end{ltxfigure}

These results show that we can fit all experimental data for $p+p$ and $A+A$, for $\sqrt{s}$ ranging from 0.2 TeV up to 7 TeV, and for different centralities, with a fixed temperature, $T$, and using only $q$ as free parameter. Also, the new procedure allows us to study $q$ as a function of $\sqrt{s}$ or as a function of the centrality for different values of $T$. In Fig.~\ref{fig:AllDatas_qSeveralT-a} we show $q$ {\it vs} $\sqrt{s}$ for $p+p$ collisions,  and in Figs.~\ref{fig:AllDatas_qSeveralT-b}~-~\ref{fig:AllDatas_qSeveralT-d} we show $q$ {\it vs} centrality for $A+A$ collisions, as obtained for different values of $T$. 

In Fig.~\ref{fig:AllDatas_qSeveralT-a} we observe that $q$ increases monotonically with $\sqrt{s}$, the shape being approximately described by a sigmoidal function of the temperature. A sigmoidal behaviour was already conjectured by C. Beck~\cite{Beck}, where it was shown that we need $q \le $ 1.2 to ensure integrability of thermodynamic functions. Here we show that experimental data leads to a value which is below that limit, and therefore a physical interpretation of the results is possible.

In Figs. \ref{fig:AllDatas_qSeveralT-b} - \ref{fig:AllDatas_qSeveralT-d} we show $q$ as a function of the centrality for all nucleus-nucleus collisions studied here, and for all values of $T$. For comparison we show values of $q$ found in the Literature, and we see that they fall in the same range obtained here. In all cases $q$ presents very small variations in the entire range of centrality, and it is approximately constantly, mainly when the highest values for $T$ are used. For low $T$, the entropic factor presents an approximately linear behaviour.


      \begin{ltxfigure}[!h]
\setcaptype{figure} 
       \centering
       \subfigure[p+p  collision.]
                   {\label{fig:AllDatas_qSeveralT-a}\includegraphics[width=7.0cm]{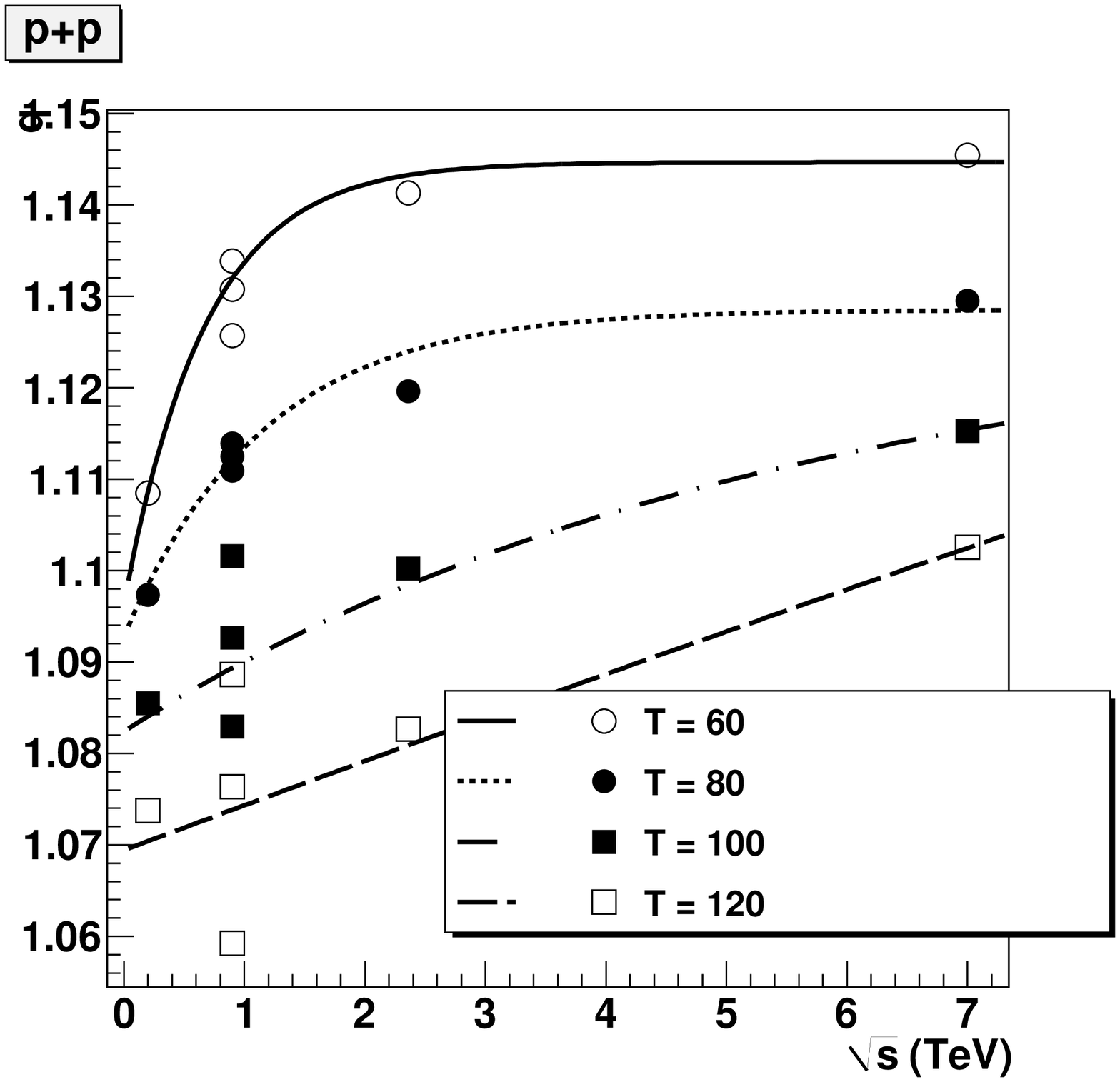}}
       \subfigure[Au+Au  collision at $\sqrt{s_{NN}} = 62,4$ GeV.]
                   {\label{fig:AllDatas_qSeveralT-b}\includegraphics[width=7.0cm]{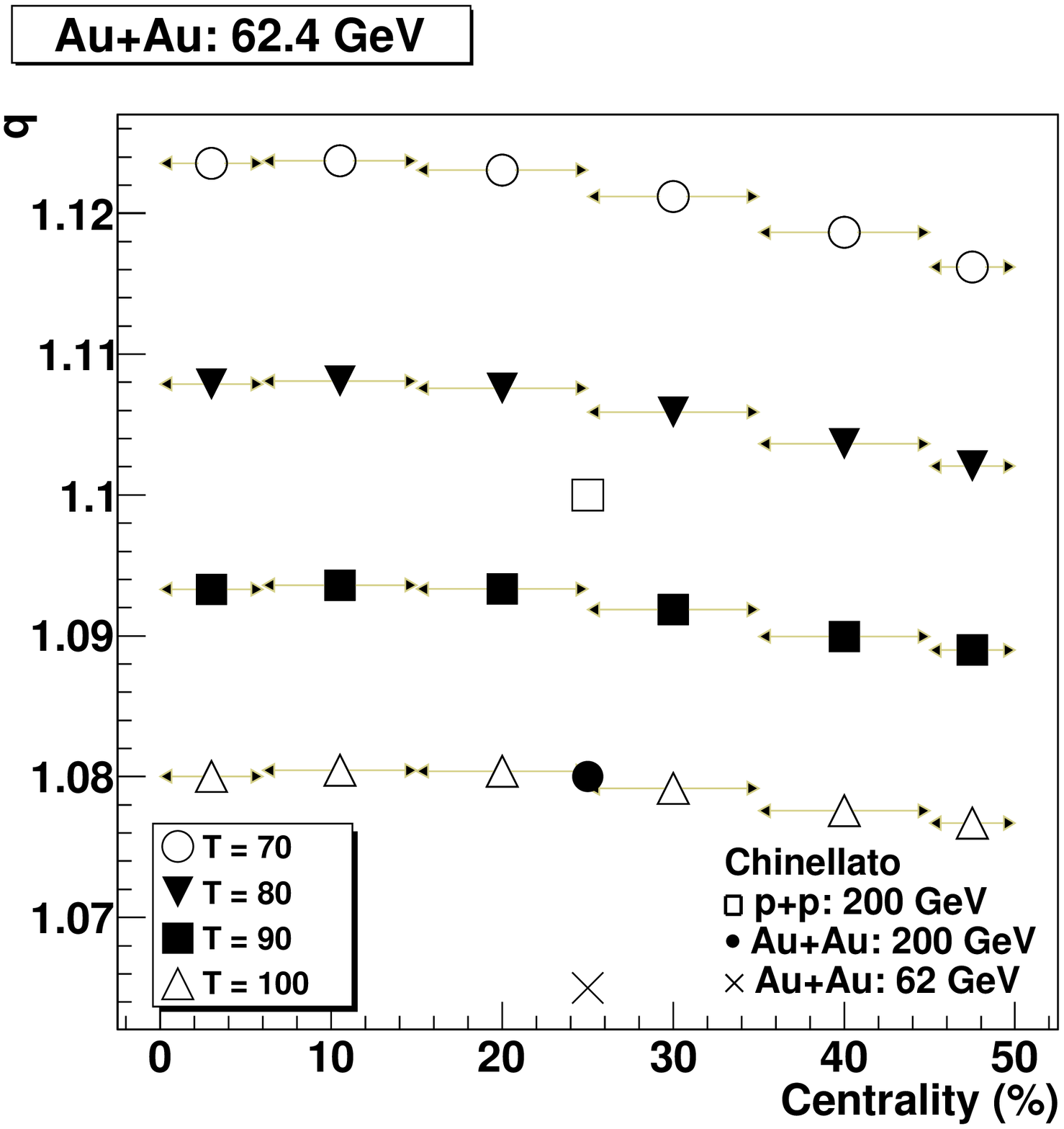}}
       \quad
       \subfigure[Au+Au  collision at $\sqrt{s_{NN}} = 200$ GeV.]
                   {\label{fig:AllDatas_qSeveralT-c}\includegraphics[width=7.0cm]{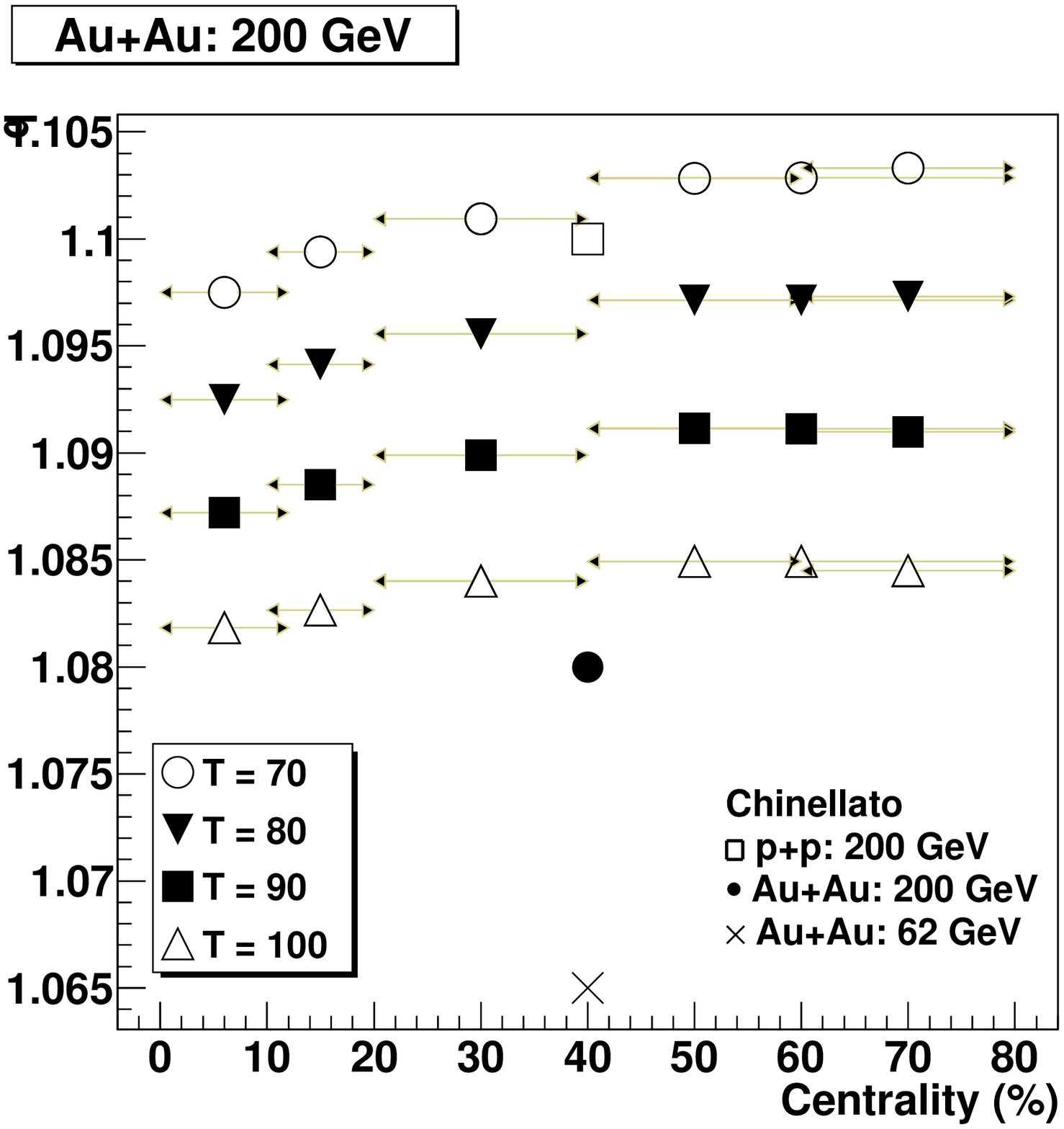}}
       \subfigure[Cu+Cu  collision at $\sqrt{s_{NN}} = 200$ GeV.]
                   {\label{fig:AllDatas_qSeveralT-d}\includegraphics[width=7.0cm]{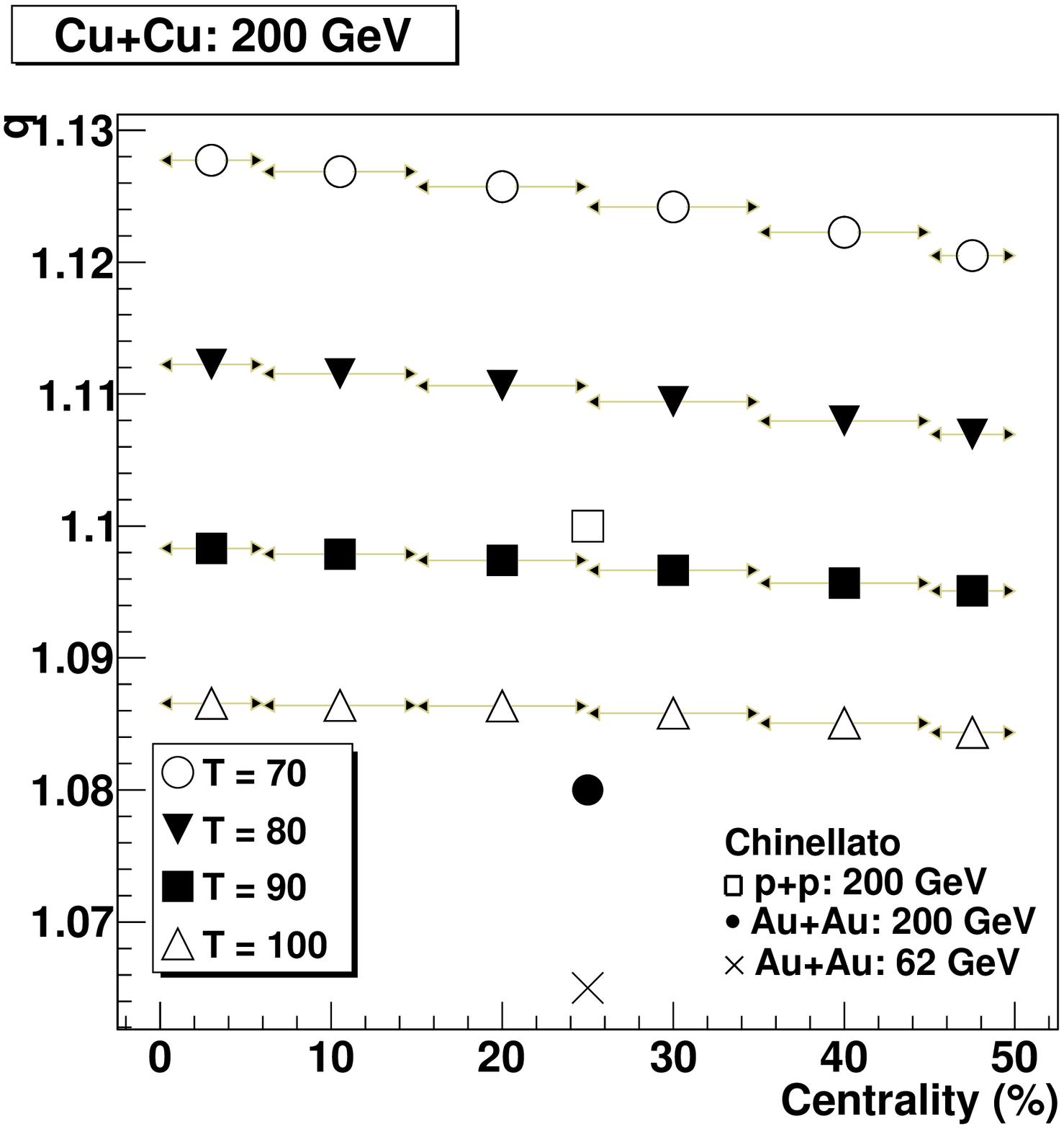}}
          \caption{The values for $q$ corresponding to the best fit with fixed $T$, for different values of temperature. Full symbols correspond to values from~\cite{Chinellato2010}. The lines in (a) are the best fitted curves for a sigmoidal function to the data.}
       \label{fig:AllDatas_qSeveralT}
      \end{ltxfigure}

The results obtained here are in agreement with an analysis performed in~\cite{Cleyman_Worku}, where the $p_t$-distributions for different particles produced in $p+p$ and $A+A$ collisions at center-of-mass energy of 0.9 TeV were studied, resulting in $T$ and $q$ constant for all particles with $T\approx$75 MeV and $q\approx$1.15. The formula used in~\cite{Cleyman_Worku} is slightly different from Equation~\ref{pt} used here, because it is restricted to particles of the same mass, and therefore cannot be used to study general secondary distributions as we do here. Also they use an occupation number formula which is consistent with the thermodynamic relations but, as discussed in~\cite{Deppman}, the differences are not significant for the analysis performed here.

In conclusion, the study presented here gives strong evidences for a limiting temperature for the hadronic system formed in hadron-hadron collisions. Also, it allows observing that $q$ increases monotonically with the center-of-mass energy, with a behaviour that is approximately sigmoidal. For nucleus-nucleus collisions, $q$ is approximately constant for all centralities, or show a small linear variation, depending on the temperature adopted.

This work received support from the Brazilian agency, CNPq, under grant 305639/2010-2 .

\bibliographystyle{aipproc}

\end{document}